\documentstyle[prd,aps,epsfig,eqsecnum,amssymb,amsmath,floats]{revtex}

\def\abstract{\par
\ifpreprintsty %
\vskip2.5pc
\begin{center}%
{\large \abstractname\par}%
\end{center}%
\vskip.5pc
\fi
\bgroup
\ifdim\prevdepth=-1000pt \prevdepth0pt\fi
\hsize2\columnwidth
\if@twocolumn\else\leftskip=0.10753\textwidth \rightskip\leftskip\fi
\dimen0=-\prevdepth \advance\dimen0 by17.5pt \nointerlineskip
\small\vrule width 0pt height\dimen0 \relax
}

\begin{document}

\title{Baryon asymmetry of the Universe from evaporation of primordial black holes.}
\author{
E. V. Bugaev, M. G. Elbakidze and K. V. Konishchev
}
\address{
Institute for Nuclear Research, Russian Academy of Sciences, 
Moscow 117312, Russia
}
\date{
\today
}

\twocolumn[
\begin{@twocolumnfalse}

\maketitle
\begin{abstract}
The process of baryogenesis through the evaporation of black holes formed at the end of inflation phase is considered.
The increase of black hole mass due to accretion from the surrounding radiation after the reheating is taken into account. 
It is shown that the influence of the accretion on the baryogenesis is important only  in the case when the  initial values of black hole mass are larger than
$\sim 10^{4}\mbox{ g}$. The behavior of calculated baryon asymmetry, as a function of model parameters, is studied.
\end{abstract}
\vskip12.5pt
\end{@twocolumnfalse}
]

\section{INTRODUCTION}

First discussions about possible connection between baryon asymmetry and primordial black hole evaporations 
appeared at the middle of seventies, just after discovery of a phenomenon of
the black hole evaporation. The  possibility of an appearing  of the excess  of  baryons over  antibaryons
in the process of  evaporation of  primordial black holes was noticed  in works \cite{10,11}. The mechanism  discussed in 
\cite{10,11} doesn't require the  non-conservation of  baryon number in an underlying microscopic theory.
Detailed calculations using this mechanism  were done  in  works \cite{12}.

Later,  this question was studied  \cite{1,2} in the context of 
grand unified theories (GUTs) (i.e., theories in which baryon number is not conserved). The 
idea is simple: when black holes decay by the emission of Hawking radiation, they may 
emit baryon-number-violating Higgs particles (and/or leptoquarks) whose decays naturally generate baryon asymmetry.

In the work of J.Barrow et al.~\cite{3} baryogenesis via primordial black holes was considered 
using GUT and extended inflation scenario. The formation of very light primordial black holes 
(which disappear, as a result of evaporation, before nucleosynthesis, without any trace, except 
of the net baryon asymmetry) seemed to be most probable in inflationary models with first 
order phase transition. Recently, however, it was shown that in some modern variants of second 
order inflation models \cite{4,5} the formation of small black holes (and, consequently, baryon 
asymmetry production via black hole evaporation) is also quite possible.

In two recent works \cite{6,7} the baryogenesis through the evaporation of primordial black 
holes was studied quantitatively. Authors of these works argued that this scenario of baryogenesis 
can, in principle, explain the observed baryon number of the Universe (which is constrained by 
primordial nucleosynthesis data to be in range $(1.55-8.1)\cdot 10^{-11}$). However, the distinct 
conclusion of works \cite{6,7} is that a sufficient number of black holes can survive beyond 
the electroweak phase transition and therefore, the baryon excess produced is not washed out 
by sphaleron transitions. These conclusions are in contradiction with qualitative statements 
of pioneering work of J.Barrow et al. \cite{3}. Authors of \cite{6,7} claim that the difference 
between their result and those of J.Barrow et al. can be explained by taking into account the
accretion (accretion term was omitted in formulas of work \cite{3}).

Aside from the problem of the possible erasing of produced baryon asymmetry by sphaleron transitions,
it is important to have the definite answer on the following question: is it possible to obtain by such mechanism
(i.e., by primordial black hole evaporations) the cosmologically interesting value of baryon asymmetry?

In present work we try to answer just this question. We  calculate the baryon asymmetry using the same assumptions, as in the work of J.Barrow et al. 
The main attention is paid to rigorous solution of kinetic equations (containing accretion term) and 
to comparison of exact results with predictions based on approximate formulas derived in \cite{3}.

\section{BASIC ASSUMPTIONS
}
\label{sec:sec1}
1. A generation of the black holes in early Universe may take place in many  inflationary 
scenarios. The main assumption is that  in the early Universe there was the period of inflationary expansion
and the inflation is completed by a symmetry-breaking phase transition. The 
energy scale of symmetry breaking is $\sigma_0\sim 10^{16}\mbox{ GeV}$ (GUT scale).

Clearly, the concrete mechanisms of black hole production are not the same  in different inflationary models.

1a. Inflation of "old" type (extended inflation, first order inflation):
the Universe exits from a false-vacuum state by bubble nucleation.
Reheating and thermalization of the Universe proceeds through bubble
collisions at the end of inflation \cite{2}.
False vacuum energy density is
\begin{equation}
\rho_{v}\sim \xi \sigma_0^4\;\;\;;\;\;\;\xi\sim 10^{-4}\;\;\;,
\end{equation}
so
\begin{equation}
\rho_v\sim 10^{60}\mbox{ GeV}^4\equiv M^4.
\end{equation}
Hubble parameter at the end of inflation is
\begin{equation}
H_{end}=\sqrt{\frac{8\pi}{3}\cdot\frac{\rho_v}{m_{pl}^2}}\sim 10^{12}\mbox{ GeV}\;\;\;,
\end{equation}
so the time of the end of inflation is
\begin{equation}
t_{end}\sim \frac{1}{H_{end}}\sim 10^{-35}\mbox{ s}.
\end{equation}
Usually one supposes that
\begin{equation}
t_{RH}\approx t_{end}\approx t_f\ll \tau_h 
\end{equation}
($t_{RH}$ is the time  of reheating, $t_f$ is the time of formation of black holes, $\tau_h$ is life-time of black hole).

Black hole production proceeds 1) via the gravitational instability
of inhomogeneities formed during the thermalization phase
(i.e., during the bubble wall collisions) or 2) via the appearance 
of trapped regions of false vacuum caught between bubbles of
true vacuum.

1b. "New" or "slow-roll-over" inflation:
in this case black holes are produced through collapses of 
the overdense regions in space. So, for large probability of
primordial black hole production one must exist large amplitudes of primordial
density fluctuations at small scales.

Such fluctuations appear during inflation. The overdense region
of mass $M$ can produce black hole when this fluctuation crosses
horizon inside. At this time $M$ is equal to horizon mass $M_h$,
and black hole produced has mass $M_{BH}$ which is close to $M_h$,
\begin{equation}
M_{BH}\sim 0.1 M_h\;\;\;.
\end{equation}
The time of PBH formation is 
\begin{equation}
t_f\sim \frac{8 M_h}{m_{pl}}\sim 10^2\frac{M_{BH}}{m_{pl}}t_{pl}.
\end{equation}
If, e.g., $M_{BH}$ is about $10^3\mbox{ g}$, then
\begin{equation}
t_f \sim 10^{-33}\mbox{ s}. 
\end{equation}
At this time, the corresponding scale factor is
\begin{equation}
a_f\sim (H_0^{3/2}(2.4\cdot 10^4)^{-3/4}t_f^{3/2})^{1/3}\sim 10^{-27}
\end{equation}
and the comoving length scale of the perturbation is
\begin{equation}
\lambda_f\sim \frac{c t_f}{a_f}\sim 10^{-14}\mbox{ pc}.
\end{equation}
Further, we use the known formula
\begin{equation}
N_{\lambda_f}=45+\ln\frac{\lambda_f}{1\mbox{Mpc}}+\frac{2}{3}
\ln\frac{M}{10^{14}\mbox{GeV}}+\frac{1}{3}\ln\frac{T_{RH}}{10^{10}\mbox{GeV}}.
\end{equation}
Here, $M\sim 10^{15}\mbox{ GeV}$ (it is $\sqrt[4]{\rho_v}$), $T_{RH}$ is
reheating temperature; $N_f$ is a number of e-folds before the end of
inflation beginning from the moment when the scale crosses horizon outside.

If, e.g., $T_{RH}\sim 10^{11}\mbox{ GeV}$, one has $N_{\lambda_f}\sim 1$.
Now it is clear that the fluctuations responsible for $M_{BH}\sim 10^3\mbox{ g}$
are formed just near the end of inflation. So, in this sense, black holes with 
$M_{BH}\sim 10^2-10^3\mbox{ g}$ are lightest ones.

The large amplitudes of density perturbations at small scales corresponding
$N_{\lambda_f}\sim 1$ are naturally obtained in the hybrid inflation model of refs.\cite{4,5}.
 
If $T_{RH}\sim 10^{11}\mbox{ GeV}$, the reheating time is
\begin{equation}
t_{RH}=0.3\frac{1}{\sqrt{g_*}}\frac{m_{pl}}{T_{RH}^2}\sim 10^{-29}\mbox{ s}.
\end{equation}
So, black holes can be produced even at a time before reheating, when the Universe is dominated
by the oscillations of the inflation field.

2. We assume that all produced primordial black holes have the same mass. It will be approximately so
in hybrid inflation-type models, where the sharp maximum of density fluctuation amplitude exists at
some definite scale.

3. We assume that at $t_{RH}$ the part of energy density is in black holes:
\begin{equation}
\rho(t_{RH})=\beta\rho_{BH}(t_{RH})+(1-\beta)\rho_R(t_{RH}).
\end{equation}
$\beta$ is the free parameter of our model.

\begin{figure}[t!]
\label{fig:fig1}
\epsfig{file=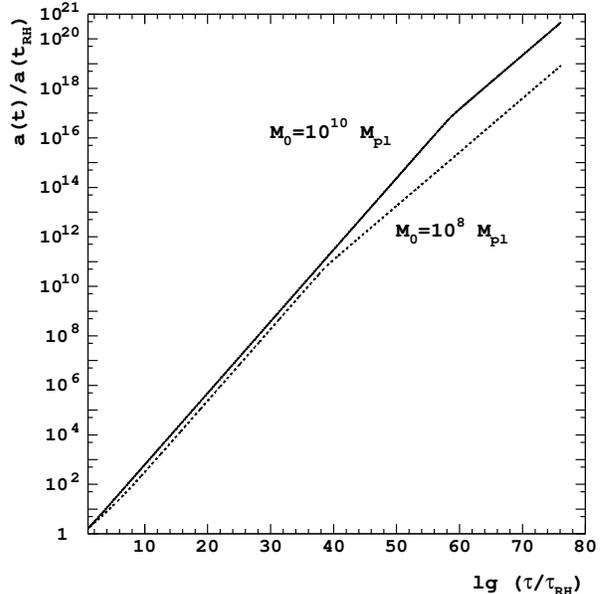,width=\columnwidth}
\caption{The time evolution of scale factor parameter $\alpha$, for $\beta=0.1$ and for two values of $M_0$.}
\end{figure}

In the following we will consider only black holes having the life-time $\tau_h$, which is much larger
than $t_{RH}$,
\begin{equation}
\tau_h\gg t_{RH}.
\end{equation}

We will see that the most interesting predictions for baryon asymmetry don't depend
on $t_{RH}$ and are the same for both inflation scenarios provided we use in both cases
the parameter $\beta$.

\section{KINETIC EQUATIONS}

\begin{figure}[t!]
\label{fig:fig2}
\epsfig{file=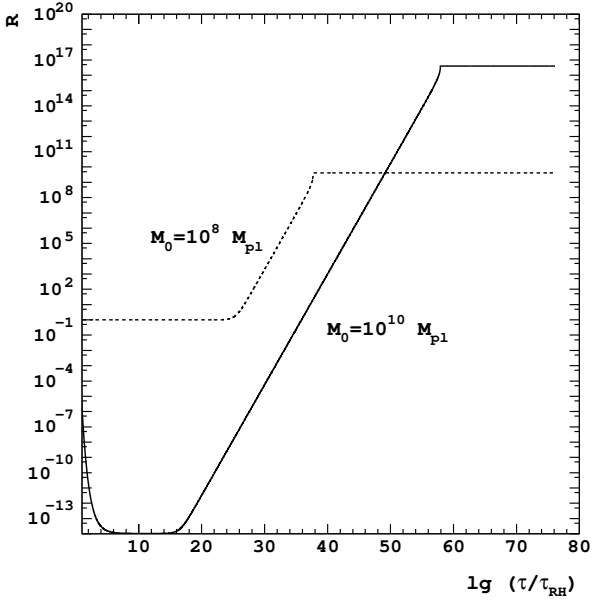,width=\columnwidth}
\caption{The dependence of function  $R$ on $\tau$, for $\beta =0.1$ and for two values of $M_0$.}
\end{figure}

Evolution of black hole mass is described by the  equation
\begin{equation}
\label{21}
\dot M_{BH}=A M_{HB}^2 -\frac{\alpha (M_{BH})}{M_{BH}^2}.
\end{equation}
Accretion term $A M_{BH}^2$ is equal to $\sigma_{abs}\rho_R c$, where $\sigma_{abs}$ is  
cross section of absorbtion  of relativistic particles by a black hole,
\begin{equation}
\sigma_{abs}=27\pi M_{BH}^2\cdot \frac{G^2}{c^4},
\end{equation}
$\rho_R$ is an energy density of the radiation.
Evaporation term is $-\frac{\alpha (m)}{m^2}$, where $\alpha (m)$ counts the degrees
of freedom of the black hole radiation ($m$ is an instantaneous value of the black hole mass).

In the following we will use the value \cite{8,9}
\begin{equation}
\alpha (m) = const = 80\cdot 10^{25}\;\frac{\mbox{g}^3}{\mbox{s}}.
\end{equation}

\begin{figure}[t!]
\label{fig:fig3}
\epsfig{file=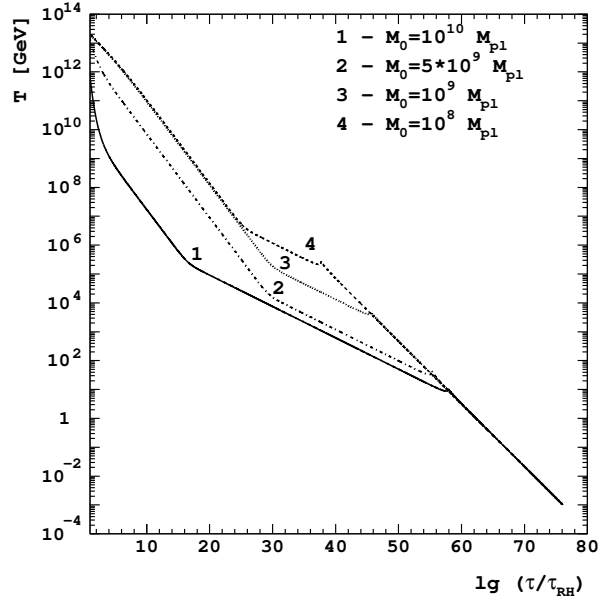,width=\columnwidth}
\caption{The $\tau$ - dependence of the temperature of radiation in the 
Universe, for $\beta =0.1$ and for several values of $M_0$.}
\end{figure}

\begin{figure}[t!]
\label{fig:fig4}
\epsfig{file=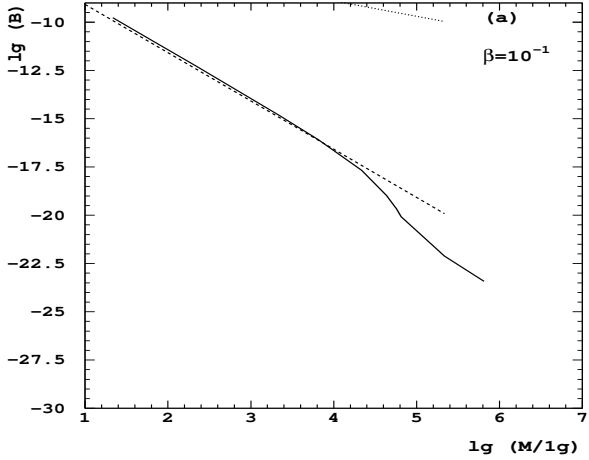,width=\columnwidth,height=0.78\columnwidth}
\epsfig{file=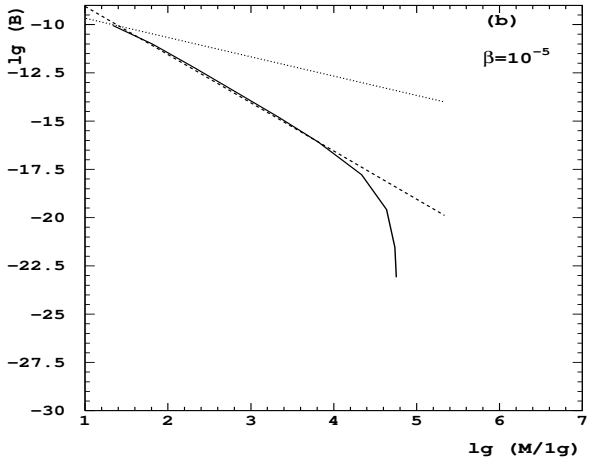,width=\columnwidth,height=0.78\columnwidth}
\epsfig{file=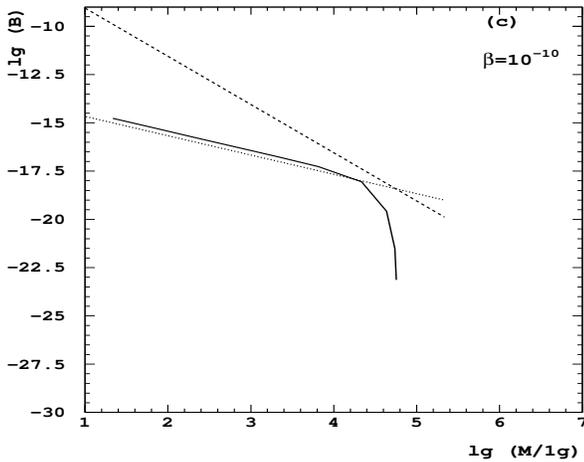,width=\columnwidth,height=0.78\columnwidth}
\caption{Baryon asymmetry parameters as a function of $M_0$ for three values of $\beta$ (solid lines).
Dashed lines: calculation by approximate formula (\ref{34}). Dotted lines: calculation by the formula (\ref{34})
with substitution (\ref{sub}).
}
\end{figure}

Evolution of the radiation energy density is given by the equation
\begin{equation}
\label{24}
\dot \rho_R=-4\frac{\dot a}{a}\rho_R -{\dot M_{BH}}n_{BH},
\end{equation}
\begin{equation}
n_{BH}(t)=\frac{\rho_{BH}(t_{RH})}{M_0}\cdot\frac{a^3 (t_{RH})}{a^3 (t)},
\end{equation}
\begin{equation}
\rho_{BH}(t_{RH})=\beta \rho(t_{RH}).
\end{equation}
Here,  $n_{BH}$ is number density of primordial black holes, $M_0$ is initial value of black hole mass, $a(t)$ is scale factor.

Evolution of scale factor
 is given by Friedmann - Einstein equation:
\begin{equation}
\label{27}
\left(\frac{\dot a}{a}\right)^2=\frac{8\pi G}{3}(\rho_R+M_{BH}\cdot n_{BH})
.
\end{equation}
In this paper we use, for simplicity, the Einstein - de~Sitter model.

Equation for evolution of the  baryon number is
\begin{equation}
\label{28}
\dot n_{B} = -\frac{\dot M_{BH} c^2}{3 k T_{BH}}\cdot n_{BH}
\cdot \epsilon f_H - 3\frac{\dot a}{a} n_{B}.                
\end{equation}
We assume here, approximately, that for $d M_{BH}$ change of black hole mass 
we have $d M_{BH}\cdot c^2/3kT_{BH}$ particles produced ($T_{BH}$ is temperature of 
black hole connected with $M_{BH}$ by Hawking relation $kT_{BH}=\frac{\hbar c^3}{8\pi G M_{BH}}$);
$f_H$ is the fraction of $X$ particles decaying with violation of B.
Typically, $f_H\sim \frac{1}{g_*}\sim 10^{-2}$.
Baryon number is obtained by introducing the factor $\varepsilon$ which is given by the relation
\begin{equation}
\varepsilon\equiv \sum\limits^{}_i B_i \frac{\Gamma (X\to f_i) - \Gamma (\bar X \to \bar f_i)}{\Gamma_X}
.
\end{equation}
$\varepsilon \ne 0$ if C,CP are not conserved.
For obtaining baryon asymmetry one must divide $n_B$ on entropy $s$ which is given by
\begin{equation}
s(t)=\frac{2\pi^2}{45}g_* T^3 (t)\left(\frac{k}{\hbar c}\right)^3,
\end{equation}
\begin{equation}
T\sim \rho_R^{1/4}.
\end{equation}
It is very convenient \cite{3,6} to use dimensionless variables:
$$
m=\frac{M_{BH}}{M_{0}}\;\;\;;\;\;\; \alpha\equiv\frac{a(t)}{a(t_{RH})}\;\;\;;\;\;\;
R=\frac{\rho_R\alpha^4}{(1-\beta)\rho(t_{RH})}
,
$$
\begin{equation}
\end{equation}
$$
\tau=\sqrt{G\rho(t_{RH})}t,
$$
for which we have simple initial conditions
$$
m(\tau=\tau_{RH})=\alpha (\tau=\tau_{RH})=R(\tau=\tau_{RH})=1,
$$
\begin{equation}
\end{equation}
$$
\tau_{RH}=\sqrt{G\rho (t_{RH})} t_{RH}.
$$
We assume, approximately, that $X$-particles are evaporated when the following condition
holds:
\begin{equation}
3kT_{BH}=3k\frac{\hbar c^3}{8\pi G M_{BH}}\ge M_{X}c^2,
\end{equation}
\begin{equation}
M_{BH}^{thr}=\frac{3 m_{pl}^2}{8\pi M_X}.
\end{equation}
Here, $M_X$ is a mass of $X$ particles.

Finally, we have the following free parameters:
\begin{equation}
M_0\;\;\;,\;\;\;\rho (t_{RH})\;\;\;,\;\;\;\beta\;\;\;,\;\;\; M_X\;\;\;,\varepsilon .
\end{equation}

Some  results of the solution of the system of kinetic equations (\ref{21}), (\ref{24}), (\ref{27}) and (\ref{28})
are presented on Fig.1-4. All calculations are carried out with  $\rho  (t_{RH})=10^{55}\mbox{ GeV}^4$.
Such value  of $\rho (t_{RH})$ corresponds to $t_{RH}\sim 10^{-33}\mbox{ s}$ and  reheating temperature $T_{RH}\sim 10^{13}\mbox{ GeV}$.

\section{APPROXIMATE FORMULAS FOR B
}

One can easily show \cite{3} that for practically important case, when
\begin{equation}
\label{31}
\beta \gg \sqrt{\frac{t_{RH}}{\tau_h}},
\end{equation}
and when accretion is not important $(M_0\le 10^3\mbox{g})$,
there is an approximate solution of the kinetic equations for $B$ which 
is given by the formula
\begin{eqnarray}
\label{34}
B=\left(\frac{30}{g_*}\right)^{1/4}\sqrt{\pi}\left(\frac{\hbar}{c}\right)^{3/4}
\rho^{1/4} (t_{RH})\times \nonumber\\
\\
\frac{(M_{BH}^{thr})^2}{m_{pl}^2 M_0}\cdot \frac{\varepsilon}{g_*}
\cdot\left(\frac{t_{RH}}{\tau_h}\right)^{1/2}.
\nonumber
\end{eqnarray}
Using the relation
\begin{equation}
\rho(t_{RH})=\frac{3}{32}\cdot \frac{m_{pl}^2}{t_{RH}^2}\sim \frac{1}{t_{RH}^2},
\end{equation}
one can   see that $B$ doesn't depend on $\rho (t_{RH})$ or $t_{RH}$.

The  expression for $\tau$  is obtained from the equation \begin{equation}
\dot M_{BH}=  - \frac{\alpha}{M_{BH}^2}
\end{equation}
and is given by
\begin{equation}
\tau_h= \frac{M_0^3}{3\alpha}\;\;\;.
\end{equation}
For $M_0\sim 10^2-10^3 \mbox{ g}$ one has
\begin{equation}
\tau_h\sim 10^{-21}-10^{-18}\mbox{ s}
.
\end{equation}
In our  numerical calculations we used $t_{RH}=10^{-33}\mbox{ s}$.  
So one has
\begin{equation}
t_{RH}\ll \tau_h\;\;\;.
\end{equation}

From here, one  has the condition for parameter $\beta$:
\begin{equation}
\label{310}
\beta\gg\sqrt{\frac{t_{RH}}{\tau_h}}\sim(10^{-6}-10^{-7}).
\end{equation}
For  such values of $\beta$, $t_{RH}$, $\tau_h$ one has the final formula:
\begin{eqnarray}
\label{aaa}
B\cong 10^{-14}\cdot \left(\frac{M_0}{10^3\mbox{ g}}\right)^{-5/2}\times\nonumber\\
\\
\left(\frac{M_X}{10^{14}\mbox{GeV}}\right)^{-2}\cdot\left(\frac{\varepsilon}{1}\right).\nonumber
\end{eqnarray}

Condition (\ref{31}) means  \cite{3} that the black hole  energy density dominates at the time of evaporation. 
If the value of parameter $\beta$ is such that 
\begin{equation}
\beta \ll \sqrt{\frac{t_{RH}}{\tau_h}}\;\;\;,
\end{equation}
the evaporation occurs while background radiation dominates the energy density of the Universe (with the entropy arising 
mainly from the radiation). In this case  the approximate formula for $B$  \cite{3} can be obtained from Eq.(\ref{34}) by substitution
of last factor:
\begin{equation}
\label{sub}
\left(\frac{t_{RH}}{\tau_h}\right)^{1/2}\to\frac{\beta}{(1-\beta)^{3/4}}.
\end{equation}

\section{conclusion}

The main conclusion of our work is the following: the predicted baryon asymmetry in  the region $M_0\alt 10^4\mbox{ g}$
is well described by the approximate formulas (\ref{34}) and (\ref{34}) with substitution of (\ref{sub}).   
These formulas were obtained without taking into account
the  accretion, and it means that the accretion process is not important for small values of initial black hole masses.
At  larger masses  accretion becomes to be important  (the same was argued in the work \cite{7}). One can see 
from Fig.4 that the accretion leads to a significant  decrease of $B$ (this decrease
is  a consequence of a significant growth of entropy after an evaporation of black  holes enlarged due to accretion).

The  resulting formula for $B$ (Eq.\ref{34}) doesn't depend on $\rho (t_{RH})$ and on $t_{RH}$ and, therefore,  is valid for
both variants of inflationary scenario mentioned in Sec.\ref{sec:sec1} (as far   as the condition (\ref{31}) is fulfilled).
It is seen that, if $M_0\sim  10^3\mbox{ g}$ and $M_X\sim 10^{14}\mbox{GeV}$, the predicted asymmetry  is quite small  (even for
$\varepsilon\sim 1$). This our  calculation strongly disagrees with the corresponding results of work \cite{7}. The effect (baryon asymmetry)
can  be large if $M_0< 10^3\mbox{  g}$ and/or $M_X\ll 10^{14}\mbox{ GeV}$ (e.g., if $M_X\sim10^{11}\mbox{ GeV}$).

It follows from Fig.3 that the temperature of the Universe at a moment of the evaporation is smaller 
than $\sim 100\mbox{ GeV}$ only in  the case, when $M_0\agt 10^5 \mbox{ g}$. For smaller values of $M_0$ (for which the value of $B$
can be acceptably large) one has $T_{ev} > 100\mbox{ GeV}$, and the  problem connected with the sphaleron transitions 
exists.  Evidently, the baryon asymmetry produced by primordial black hole evaporations can 
survive only if nonzero $(B-L)$-value is generated in decays of Higgs particles of GUT.

\end{document}